\documentclass[11pt]{article}

\usepackage[preprint]{acl}

\usepackage{times}
\usepackage{latexsym}

\usepackage[T1]{fontenc}
\usepackage[utf8]{inputenc}
\usepackage{microtype}
\usepackage{inconsolata}
\usepackage{graphicx}
\usepackage{booktabs}
\usepackage{amsmath}

\usepackage{xcolor}

\title{Simulating Hate Speech Cascades with Multi-LLM Agents:\\
Empirical Grounding, Modeling Fidelity, and Intervention Strategies}

\author{Fan Huang \\
  Indiana University Bloomington \\
  \texttt{huangfan@acm.org}}

\begin{document}
\maketitle

\begin{figure*}[!t]
\centering
\includegraphics[width=\textwidth]{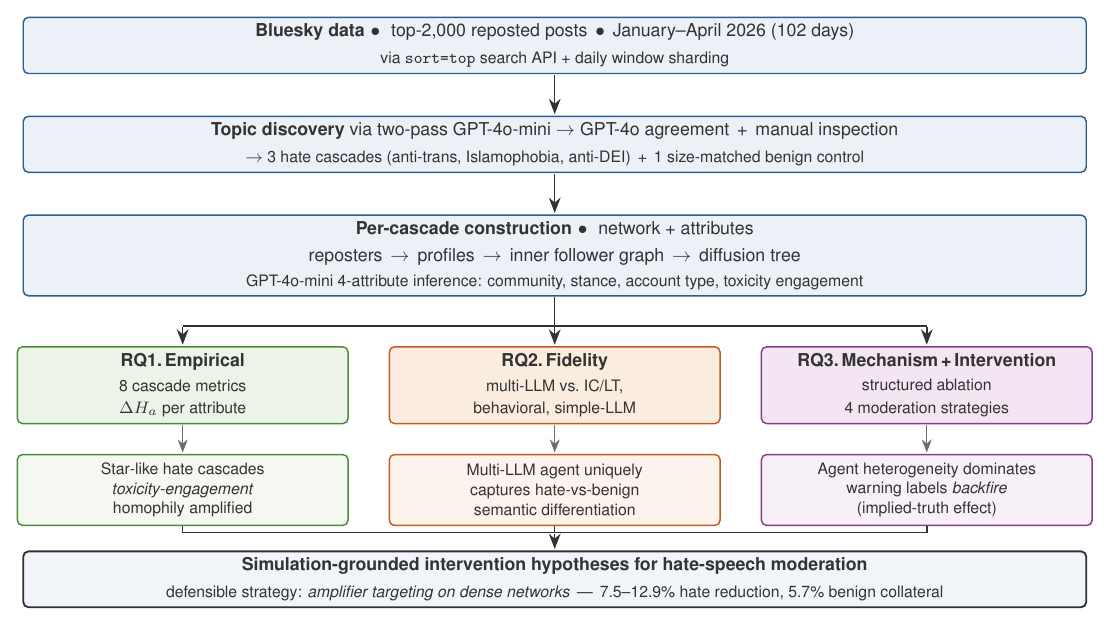}
\caption{End-to-end study pipeline. Top: Bluesky data collection (start date
January 1, 2026; end date April 12, 2026; 102-day window), two-pass
GPT-4o-mini then GPT-4o topic discovery, and per-cascade network and
attribute construction. Middle: three parallel research questions on the same
constructed cascades, each with a single-thought headline finding. Bottom:
the three tracks converge on simulation-grounded intervention hypotheses.}
\label{fig:pipeline}
\end{figure*}

\begin{abstract}
Faithful modeling of hateful-content propagation on online platforms remains
an open problem for moderation research.
Classical cascade models that do not explicitly represent the profile,
community, and content factors associated with hateful-content propagation
may yield moderation strategies that behave less effectively when deployed in
real-world scenarios. Multi-agent large language model (LLM) systems
can in principle make each reshare decision depend on the user's profile, the
surrounding community, and the post's content, but it remains unclear whether
this added flexibility actually reproduces real hateful cascades more
faithfully than classical baselines.
We study three hateful Bluesky cascades and a size-matched benign control.
In the empirical Bluesky data, we found that:
97.4--99.7\% of reposters take a hostile stance; toxicity-engagement homophily
is higher on the diffusion tree than on the follower graph for hateful
cascades%
; topology is star-like
for the hateful cascades (most reposts come directly from the root) versus
tree-like for the benign cascade (reposts propagate through multi-hop chains).
In simulation, a multi-LLM-agent simulator reproduces the stance monoculture
and the toxicity-delta direction. A structured ablation identifies agent
heterogeneity as the leading fidelity factor, and amplifier targeting on dense
networks yields 7.5--12.9\% reduction at 5.7\% benign collateral.
\end{abstract}

\section{Introduction}
\label{sec:intro}

Hate speech on social media has been linked to offline harms, from psychological
distress among targeted groups to elevated rates of racially and religiously
motivated crime \citep{williams2020hate,muller2021fanning}, and viral toxic
content reaches audiences far beyond any one community
\citep{mathew2019spread,matamoros2021racism}. Platforms therefore face a
recurring question: how does hateful content propagate through follower networks,
and which interventions dampen it without suppressing benign engagement?

A simulation account of this question must capture who reshares as a function of
profile and community context, how those decisions aggregate into cascade
structure, and how the same population responds to candidate moderation
strategies. Classical cascade models (Independent Cascade
\citep{kempe2003maximizing}, Linear Threshold \citep{granovetter1978threshold})
treat resharing as a fixed-rule probabilistic event and do not represent
these factors. Large language
models (LLMs) used as conditioning agents can in principle express them
\citep{park2023generative,argyle2023out,horton2023large,ziems2024can,gao2024large},
but it is unclear whether they reproduce real hateful cascades more faithfully
than simpler baselines.

Empirical work characterizes cascade size, depth, and virality on real platforms
\citep{vosoughi2018spread,goel2016structural}, and a hate-speech literature
documents hateful-user signatures \citep{ribeiro2018characterizing}, echo-chamber
amplification \citep{sasahara2021social}, cascade heavy tails
\citep{mathew2019spread}, ban effects \citep{chandrasekharan2017you}, and
offline-event coupling \citep{olteanu2018effect}. Warning-label studies report
both intended reductions \citep{mena2020cleaning,clayton2020real} and
implied-truth backfire \citep{pennycook2020implied}. A parallel strand evaluates
LLM agents as a social-simulation methodology
\citep{park2023generative,tornberg2023simulating,bail2024can,gao2024large,ziems2024can}.

These literatures are largely disjoint: empirical cascade studies rarely evaluate
generative simulators on the same observed networks; LLM-agent papers seldom
benchmark against classical cascade baselines on a hate-speech task; and
moderation strategies are rarely tested under a mechanism-aware model of agent
response. It therefore remains open whether multi-LLM-agent simulation provides
a measurable fidelity gain, what mechanisms drive any gain, and which
interventions are supported once mechanism is accounted for.

We organize the study around three research questions. RQ1: what structural,
temporal, and community-level regularities characterize real-world hateful
cascades, and how do they differ from a size-matched benign control? RQ2: to
what extent does a multi-LLM-agent simulator reproduce these regularities
relative to classical diffusion and simpler LLM baselines on the same networks?
RQ3: which agent-level mechanisms account for fidelity differences, and which
moderation strategies are supported by simulation-grounded counterfactuals?

We contribute
(1) a per-cascade empirical characterization of three Bluesky hateful cascades
(January--April 2026) and a size-matched benign control, with bootstrap intervals
reported per cascade rather than aggregated;
(2) a fidelity comparison of a multi-LLM-agent simulator against classical
diffusion, behavioral, and simpler LLM baselines on the same observed networks,
in which the multi-LLM agent is the only tested family that separates hateful
from benign content under a fixed population;
(3) a structured ablation in which agent heterogeneity has the largest
toxicity-delta shift among the conditions tested;
(4) simulation-based counterfactual testing of four moderation strategies, in
which warning labels enlarge hateful cascades in our simulations (consistent
with the implied-truth effect) and amplifier targeting on dense networks shows
the most favorable hateful-reduction to benign-collateral trade-off among the
four; and
(5) a prompt-framing observation that probability-prediction reduces the
persona-alignment refusals we observed with role-play prompts on the
RLHF-aligned backbones we tested. Figure~\ref{fig:pipeline} summarizes the
pipeline.

\section{Related Work}
\label{sec:background}

\paragraph{Classical cascade models.}
Information diffusion on networks is commonly modeled with the Independent Cascade (IC)
model \citep{kempe2003maximizing}, in which each edge $(u,v)$ activates independently
with probability $p_{uv}$, and the Linear Threshold (LT) model
\citep{granovetter1978threshold}, in which a node activates when the weighted fraction
of active neighbors exceeds a node-specific threshold. Epidemic-style SIR/SIS models
have been used as a related abstraction for information spread
\citep{pastor2015epidemic}. These approaches are interpretable and scalable, but they
collapse content, profile, and community context into a single transmission probability,
which limits their ability to express content-conditioned dynamics relevant to
hate-speech diffusion.

\paragraph{Empirical online cascades.}
Empirical studies characterize size, depth, and structural virality of online cascades
\citep{vosoughi2018spread,goel2016structural}, document how algorithmic ranking shapes
ideological exposure on social platforms \citep{bakshy2015exposure}, and examine
experimental spread of behavior in observable networks \citep{centola2010spread}.
We follow this
measurement style and extend it to hate-speech content with a matched benign control on
the same platform.

\paragraph{Hate-speech dynamics and moderation.}
Earlier work documents detection methods and dataset gaps
\citep{davidson2017automated,fortuna2018survey,vidgen2020directions}, hateful-user
signatures on Twitter \citep{ribeiro2018characterizing}, the role of echo chambers
\citep{sasahara2021social}, the heavy-tailed spread of hate-speech cascades
\citep{mathew2019spread}, platform-level effects of community bans
\citep{chandrasekharan2017you}, links between online hate exposure and offline
crime \citep{muller2021fanning,williams2020hate}, the way offline events
influence online hate \citep{olteanu2018effect}, and broader systematic reviews
of how racism circulates across major social-media platforms
\citep{matamoros2021racism}. On the
moderation side, warning labels and fact-check tags reduce sharing under
some conditions \citep{mena2020cleaning,clayton2020real} but can also produce
implied-truth effects on unlabeled content \citep{pennycook2020implied}. To our
knowledge, this body of work has not been used as an empirical anchor for
multi-LLM-agent cascade simulators.

\paragraph{LLM-based agents and social simulation.}
A growing line of work uses LLMs to simulate human decisions in social, economic,
and survey settings \citep{park2023generative,argyle2023out,horton2023large,ziems2024can}
and discusses social-simulation methodology
\citep{tornberg2023simulating,bail2024can,gao2024large}. These studies suggest that
LLM agents can in principle condition on profile, content, and context.

\section{Data and Network Construction}
\label{sec:data}

\paragraph{Platform and time window.}
We collect data from Bluesky, an open, decentralized social platform whose API exposes
follower graphs and repost (reshare) traces. The collection window is January 1--April 12,
2026 (102 days), from which we retrieve the top-2{,}000 most-reposted English-language
posts via Bluesky's \texttt{sort=top} search API with daily time-window sharding.

\paragraph{Topic selection.}
Topic selection is data-driven from the cascade pool rather than keyword-targeted. All
top-2{,}000 posts are classified for explicit hate speech, implicit hate speech, and
social bias by a two-pass GPT-4o-mini then GPT-4o agreement pipeline; manual inspection
of the resulting candidate set selects three primary hateful cascades spanning three
distinct dimensions and one size-matched benign control on the same platform:
(1)~Cascade~A --- anti-trans / identity-based (2{,}267 reposters);
(2)~Cascade~B --- Islamophobia / ethnic-religious (2{,}796 reposters);
(3)~Cascade~C --- anti-DEI / social-racial (2{,}942 reposters); and
(4)~benign control --- apolitical entertainment commentary (3{,}980 reposters).
Reposter counts reflect the full collected set; cascade tree sizes in
Table~\ref{tab:cascade-summary} are slightly smaller after timestamp-based diffusion
tree inference. Per ethical convention for hate-speech research, the original
Bluesky cascade-seed handles are pseudonymized in this paper using the
single-letter aliases above (and Cascades~D--F for the secondary cascades
introduced in Appendix~\ref{sec:appendix-n6}); the handle-to-alias
map is available to reviewers on request.

\paragraph{Study design.}
The investigation follows a case-study design: each of the three hateful cascades is
characterized on the full metric suite and contrasted with the benign control. With
$N=3$ hateful cascades, we report direction and magnitude per cascade rather than
distributional generalization, and flag which findings replicate across topics. Two
network layers are used: the follower network (directed graph of follow relationships
among users in the topic-scoped collection) and the reshare network (directed graph of
repost chains forming the observed cascade tree).

\paragraph{Network construction and attribute inference.}
For each cascade we collect the reposter set, their profiles, and each reposter's follow
list filtered to other reposters and the cascade root. Diffusion trees are inferred from
the inner follower graph and per-user repost timestamps obtained via the
\texttt{com.atproto.repo.listRecords} endpoint: each reposter's inferred parent is the
most recent earlier reposter they follow, otherwise the root. No downsampling is applied;
the full reposter set per cascade (2{,}241--3{,}919 nodes) is tractable at this scale.
Each user is annotated on four attribute dimensions inferred by GPT-4o-mini ($T{=}0$)
from bio text, up to 30 recent posts, and the cascade's topical context:
(i)~\emph{community identity} (the discourse community the user belongs to);
(ii)~\emph{stance on topic} (supportive, opposed, neutral, or unclear, topic-specific
labels aliased across cascades);
(iii)~\emph{account type} (individual, organization, activist, bot, or unclear); and
(iv)~\emph{toxicity engagement} (degree of prior engagement with hateful content within
the window). Labels below a confidence threshold of $0.65$ are replaced with
\texttt{unclear}; inferred attributes are treated as noisy estimates and audited via
sign stability of downstream homophily deltas across thresholds $\{0.50,0.65,0.80\}$.

\paragraph{Homophily measurement.}
For attribute $a$ with groups $g_1,g_2,\ldots$, homophily is defined as the probability
that an edge connects same-group nodes,
$H_a=P(\text{edge connects same-group nodes}\mid a)$. We measure $H_a$ on the follower
network (structural homophily) and on the diffusion tree (behavioral homophily), and
report the homophily delta $\Delta H_a=H_a^{\text{diffusion}}-H_a^{\text{follower}}$ as
a per-cascade summary of whether resharing preferentially follows same-group ties.

\section{RQ1: Empirical Cascade Characterization}
\label{sec:rq1}

\paragraph{Implicit-hate classification of the three cascades.}
We treat the three hateful cascades (Cascade~A, Cascade~B, Cascade~C)
as belonging to the \emph{implicit / coded-language} hate-speech regime on two
grounds. \emph{Content criterion:} the three seed posts express their target
stance through indirect rhetorical framings (concern-rhetoric around gender
identity; national-security and ethnic-cultural rhetoric around Islam;
meritocracy-framed criticism of DEI) rather than through direct slurs,
explicit-violence imagery, or direct calls to harm; all three were retained as
hateful-cascade seeds only after passing the GPT-4o-mini and GPT-4o classifier
passes \emph{and} the manual-inspection review documented in
Appendix~\ref{sec:appendix-hatespeech}. \emph{Propagation criterion:} the
cascades show the dynamics expected of in-group implicit hate content --- a
near-saturated hostile-stance share with no counter-speech surge (Finding~1
below) and a dense star-like reach pattern (Finding~5 below). The $N{=}6$
direction-stability extension reported in
Appendix~\ref{sec:appendix-n6} renders this distinction empirical:
three secondary cascades whose seed posts use explicit / confrontational
framings (e.g., identity-group symbolism juxtaposed with explicit-violence
imagery on Cascade~D) produce hostile-stance shares of $0.9$--$20.4\%$ and
tree-like topologies of depth $24$--$38$, contrasting sharply with the three
originals' $97.4$--$99.7\%$ and depth $4$--$6$. We accordingly read all
quantitative claims in this section as scoped to implicit / coded-language
hateful cascades.

We characterize hateful cascades on eight metrics commonly used in cascade analysis:
cascade size, cascade ratio (size normalized by network size), depth (longest reshare
chain), breadth (direct reshares from the root), structural virality (average pairwise
distance in the cascade tree, following \citet{goel2016structural}), time to saturation
($t_{50}$, time until 50\% of final size), per-hop reshare probability, and
cross-community penetration (fraction of reposters whose inferred community differs from
the seed's). Table~\ref{tab:cascade-summary} reports the cascade-level summary;
per-cascade structural bar charts, temporal profiles, per-hop reshare curves, and
homophily comparisons are listed in Appendix~\ref{sec:appendix-rq1}.

\begin{table*}[t]
\centering
\small
\setlength{\tabcolsep}{6pt}
\begin{tabular}{@{}lcccc@{}}
\toprule
Metric & Cascade~A & Cascade~B & Cascade~C & benign \\
\midrule
Topic & anti-trans & Islamophobia & anti-DEI & entertainment \\
Size           & 2{,}241 & 2{,}714 & 2{,}929 & 3{,}919 \\
Depth          & 6 & 4 & 5 & 40 \\
Breadth        & 1{,}891 & 2{,}521 & 2{,}566 & 1{,}017 \\
Breadth/Size   & 84\% & 93\% & 88\% & 26\% \\
Virality       & 2.51 & 2.16 & 2.29 & 14.82 \\
Span (h)       & 214 & 425 & 120 & 182 \\
$t_{50}$ (h)   & 13.7 & 3.7 & 6.8 & 8.3 \\
\bottomrule
\end{tabular}
\caption{Cascade-level summary for the three hateful cascades and the benign control.
Hateful cascades are star-like (high breadth/size, shallow depth); the benign cascade
is tree-like (depth 40, low breadth).}
\label{tab:cascade-summary}
\end{table*}

\paragraph{Finding 1: stance monoculture.}
Reposters in all three hateful cascades are predominantly labeled with a
hostile/critical stance: 97.8\% (Cascade~A), 99.7\% (Cascade~B), 97.4\%
(Cascade~C). Fewer than 2\% are labeled as sympathetic/affirming per cascade, with
the remainder unclear ($<2.4\%$). The benign control contains no identity-group target
and accordingly shows 0\% hostile stance. Stance-homophily $H_a$ on hateful cascades is
therefore close to a ceiling regardless of network layer, which we treat as a saturation
effect rather than a null result.

\paragraph{Finding 2: shallow cross-community penetration.}
Cross-community penetration, defined as the fraction of labeled reposters whose
\texttt{community\_identity} differs from the seed's (modal reposter community used as
fallback when the seed label is unclear), ranges from 4.2\% to 11.8\% across hateful
cascades (Cascade~A 7.8\%; Cascade~B 4.2\%; Cascade~C 11.8\%). Hop-wise
penetration (Appendix~\ref{sec:appendix-rq1}) does not increase consistently with
depth, suggesting that hateful cascades remain largely within a single community
throughout their lifetime.

\begin{figure}[t]
\centering
\includegraphics[width=\columnwidth]{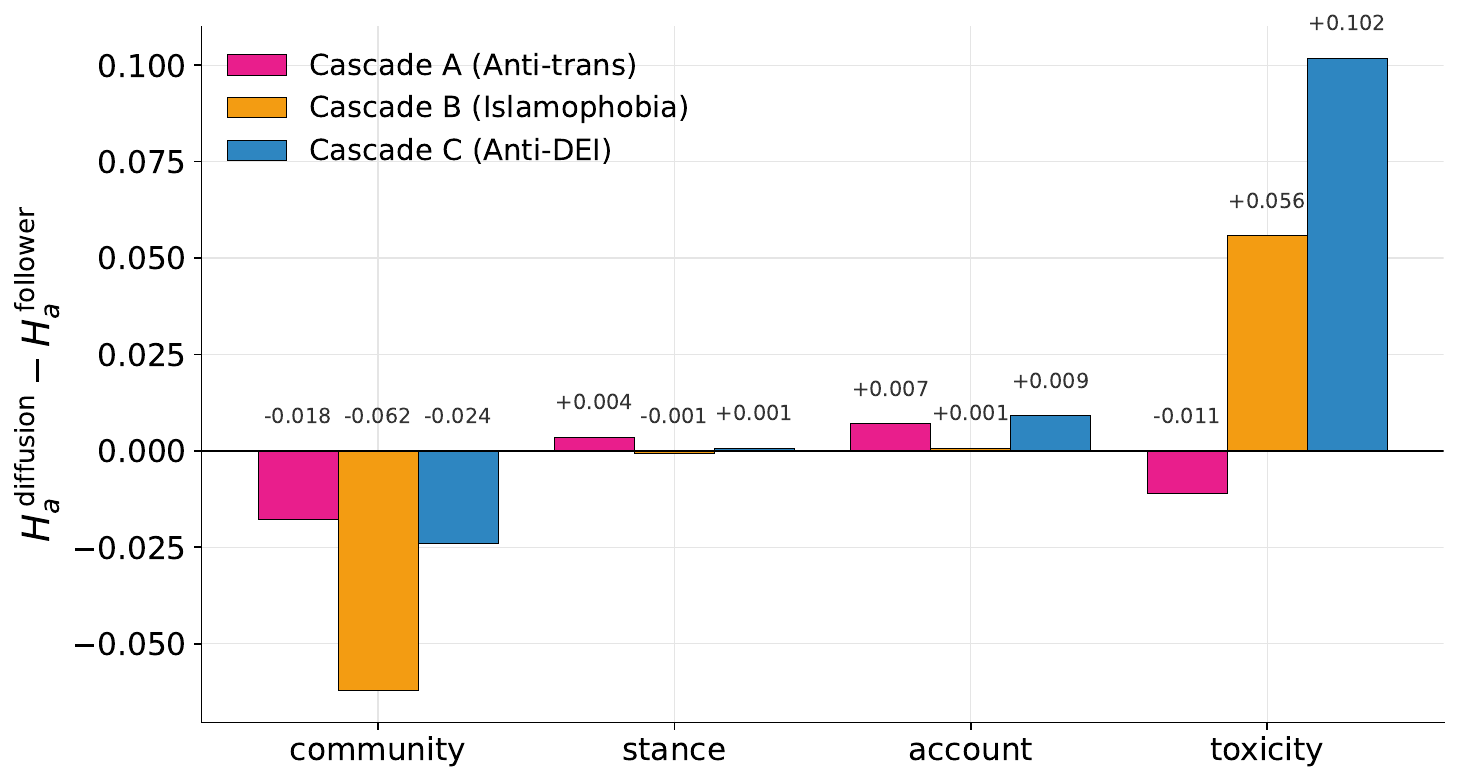}
\caption{Homophily delta $\Delta H_a = H_a^{\text{diffusion}} - H_a^{\text{follower}}$
per attribute per cascade. Negative values indicate a diffusion tree less homophilic
than the follower network; positive values, the converse.}
\label{fig:homodelta}
\end{figure}

\paragraph{Finding 3: community-identity homophily is slightly attenuated.}
For all three hateful cascades, $\Delta H_a$ on \texttt{community\_identity} is
negative (Figure~\ref{fig:homodelta}: $-0.018$, $-0.062$, $-0.024$): the diffusion tree
crosses community boundaries slightly more than the follower graph would predict. The
benign control delta is $-0.003$ (essentially zero). The Cascade~C delta has
bootstrap 95\% CI excluding zero ($[-0.044,-0.006]$) and is sign-stable across
confidence thresholds.

\paragraph{Finding 4: toxicity-engagement amplification is hate-specific.}
For two of three hateful cascades, $\Delta H_a$ on \texttt{toxicity\_engagement} is
positive: Cascade~B $+0.056$ (bootstrap 95\% CI $[+0.040,+0.071]$, threshold-stable)
and Cascade~C $+0.102$ (wide CI $[-0.097,+0.269]$ due to only 19 labeled edges).
The third (Cascade~A) shows $-0.011$, noisy under a sparse inner follower graph
(52\% isolates, 83 labeled edges). The benign control shows the opposite sign
($-0.097$).

\begin{figure*}[t]
\centering
\includegraphics[width=0.80\textwidth]{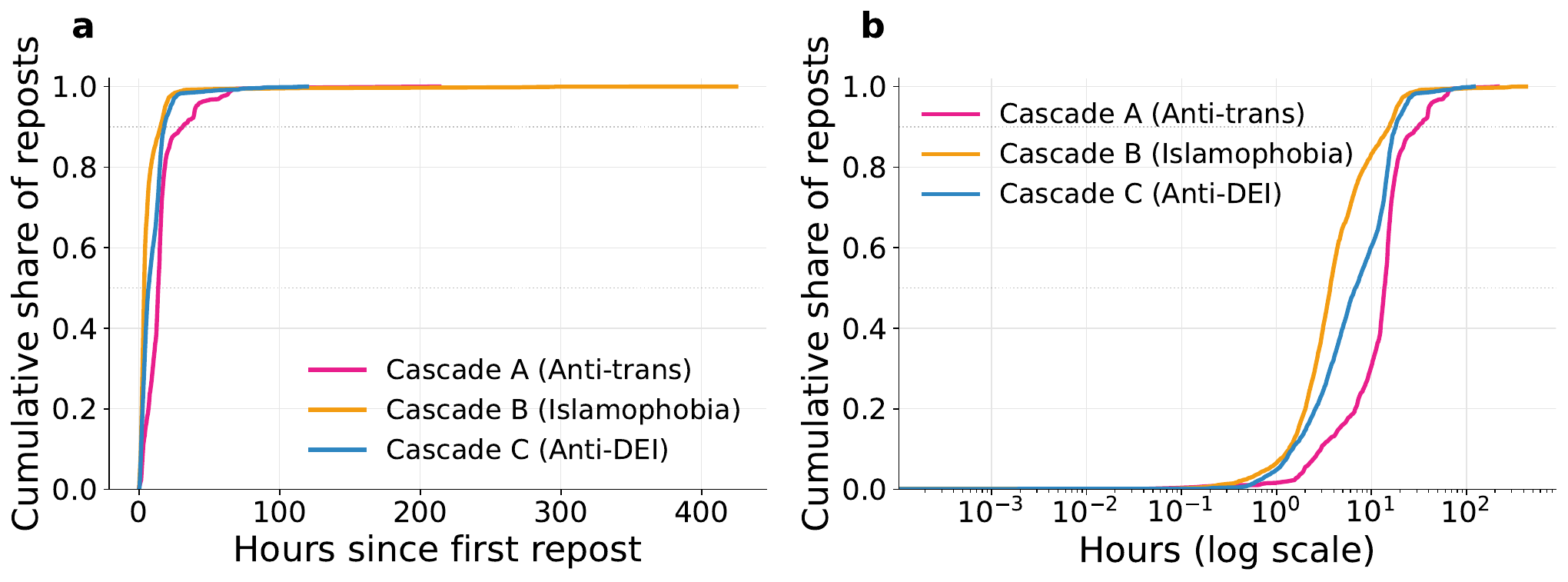}
\caption{Cumulative reshare profiles per cascade on (a)~linear and (b)~log time
axes. Cascade~B shows a fast viral burst ($t_{50}{=}3.7$\,h)
followed by a long tail spanning 17.7 days.}
\label{fig:temporal}
\end{figure*}

\paragraph{Finding 5: star-like hate, tree-like benign.}
Hateful cascades have breadth/size ratios of 84--93\% and shallow depth (4--6). The
benign control has lower breadth, depth 40, and a $6\times$ denser follower graph
($127{,}273$ versus $21{,}240$ edges for the densest hateful cascade). Combined with the
inner-graph saturation effect reported in RQ2 (Independent Cascade reaches 8--24\% of
empirical hateful cascade size on the inner follower graph and 87\% of the benign
cascade), this is consistent with hateful content propagating largely via algorithmic
feed surfaces rather than follower chains, and benign viral entertainment propagating
through follower chains. Figure~\ref{fig:temporal} shows the corresponding
cumulative reshare profiles, with Cascade~B producing a sharp viral burst
($t_{50}{=}3.7$\,h) and a long tail.

\paragraph{Robustness.}
Bootstrap 95\% intervals (1{,}000 resamples) support directional significance for the
Cascade~B toxicity-engagement delta, the Cascade~C community-identity delta,
and account-type deltas on Cascade~A and Cascade~C. All reported deltas are
sign-stable across confidence thresholds $\{0.50,0.65,0.80\}$, except stance on
Cascade~A, where the near-saturated distribution produces noise near zero. Full
bootstrap intervals and threshold tables are reported in Appendix~\ref{sec:appendix-rq1}.

\paragraph{Scope refinement from an $N{=}6$ extension.}
Three additional hateful cascades from the manual-inspection secondary pool
were collected as a direction-stability check: Cascade~D
($2{,}536$ reposters, anti-trans with explicit-violence imagery),
Cascade~E ($4{,}133$ reposters, Islamophobia), and Cascade~F
($3{,}048$ reposters, antisemitism). All four RQ1 findings replicate on
3 of 6 cascades each: the three originals pass F1, F2, and F4 while the
three secondary cascades fail; F3 (toxicity-engagement amplification) passes on
Cascade~B, Cascade~C, and Cascade~D and fails on Cascade~A,
Cascade~E, and Cascade~F. The pattern surfaces an apparent
implicit-versus-explicit hateful-cascade regime distinction: the three
secondary cascades all received substantial counter-speech responses (hostile-stance
shares of 0.9--20.4\% vs. 97.4--99.7\% on the three originals) and a tree-like
rather than star-like structure (breadth/size 17.6--44.2\% and depth
24--38 vs. 84--93\% and depth 4--6 originally). The findings reported here are
accordingly scoped to implicit / coded-language hateful cascades; the
per-cascade analysis is in Appendix~\ref{sec:appendix-n6}.

\section{RQ2: Modeling Fidelity}
\label{sec:rq2}

\paragraph{Setup.}
All simulators run on the inner follower graph constructed in
Section~\ref{sec:data}, seeded with the empirical cascade root, and are evaluated on
the same metrics as RQ1. Fidelity is reported as per-metric absolute error against the
empirical reference. We compare four model families.
(F1)~Classical diffusion: Independent Cascade (each edge activates independently with
$p_{uv}$ calibrated from empirical reshare rates) and Linear Threshold (a node
activates when the weighted sum of active neighbors exceeds a calibrated $\theta_v$).
(F2)~Behavioral heuristics: four breadth-first variants differing in per-node reshare
probability (fixed $p$; in-degree conditioned; community-similarity conditioned;
toxicity-engagement conditioned, the last directly encoding the RQ1 toxicity
mechanism).
(F3)~Simpler LLM variants: \emph{single-agent} (one shared LLM, no per-agent profile),
\emph{homogeneous-agent} (a shared generic profile across agents), and
\emph{no-network-context} (per-agent profiles but no neighborhood information).
(F4)~Multi-LLM-agent system (this work): each agent is assigned a per-user profile
(community, stance, account type, toxicity engagement) and a follower-graph
neighborhood, and uses GPT-4o-mini ($T{=}0.1$) to predict reshare probability from
profile, neighborhood context, and the post text; the probability is treated as a
Bernoulli parameter for the per-step reshare decision.

\paragraph{Prompt framing.}
A direct role-play prompt (``You are a user with this profile; would you reshare?'')
elicits safety refusals on RLHF-aligned backbones and inverts simulated hate-versus-
benign dynamics in our pilot (0--5\% amplification on hateful content and 92\% on
benign). We instead frame the agent task as behavioral prediction (``Predict the
probability $\in[0,1]$ that the user described below reshares the post'').

\paragraph{Inner-graph ceiling.}
Independent Cascade at $p{=}1$ reaches 8--24\% of empirical hateful cascade size on the
inner follower graph and 87\% of the benign cascade. All families share this ceiling,
so fidelity is reported on scale-invariant metrics (hostile-stance percentage,
toxicity-engagement homophily delta, structural virality, cross-community percentage).

\begin{table}[t]
\centering
\scriptsize
\setlength{\tabcolsep}{4pt}
\begin{tabular}{@{}lcccc@{}}
\toprule
Model & hostile\%$_{\text{err}}$ & tox $\Delta_{\text{err}}$ & viral$_{\text{err}}$ & cross\%$_{\text{err}}$ \\
\midrule
beh\_toxicity        & 0.69 & 0.014 & 1.06 & 2.01 \\
beh\_fixed           & 0.67 & 0.039 & 1.06 & 1.99 \\
beh\_degree          & 0.67 & 0.039 & 1.06 & 1.99 \\
beh\_homophily       & 0.66 & 0.048 & 1.07 & 4.40 \\
llm\_no\_context     & 1.38 & 0.059 & 0.99 & 1.43 \\
multi\_llm           & 1.00 & 0.079 & 0.73 & 1.89 \\
llm\_single          & 0.33 & 0.171 & 0.57 & 2.04 \\
llm\_homogeneous     & 0.54 & 0.213 & 0.75 & 2.50 \\
\bottomrule
\end{tabular}
\caption{Mean per-metric fidelity error across the three hateful cascades (lower is
better). The toxicity-conditioned behavioral baseline minimizes the toxicity-delta
error by encoding the RQ1 mechanism directly; the multi-LLM agent minimizes the
structural-virality error among profile-aware models and uniquely provides hateful-
versus-benign content discrimination (Finding 1 below).}
\label{tab:rq2headline}
\end{table}

\paragraph{Finding 1: content-semantic differentiation.}
With the population, network, and prompt held fixed, the multi-LLM agent produces
98--100\% hostile stance on the three hateful cascades and 0\% hostile stance on the
benign control, with the difference driven entirely by post content. Behavioral
baselines produce indistinguishable hostile-stance rates between hateful and benign
posts because they do not read content; the LLM variants without per-agent profile or
network context (\emph{llm\_single}, \emph{llm\_homogeneous}) lose the contrast in the
other direction. We read this as the principal capability behavioral baselines
structurally cannot deliver under fixed populations.

\paragraph{Finding 2: structural-virality fidelity among profile-aware models.}
The multi-LLM agent has virality error 0.73, lower than every behavioral baseline
(1.06--1.07) and the no-context LLM ablation (0.99) at matched profile fidelity. The
single-agent LLM has a lower virality error (0.57) but inflates the toxicity-delta
error to 0.171 with homogeneous reshare behavior; among profile-respecting models the
multi-LLM agent has the most favorable joint structural--content trade-off.

\paragraph{Finding 3: Cascade~A toxicity-direction.}
The multi-LLM agent is the only condition to predict a negative toxicity delta on
Cascade~A ($-0.026$ versus empirical $-0.011$). All behavioral baselines and the
remaining LLM variants assign the opposite sign. Section~\ref{sec:ablation} analyzes
the mechanism.

\paragraph{Finding 4: mechanism-specific baseline advantage.}
The toxicity-conditioned behavioral baseline minimizes the toxicity-delta error
($0.014$ versus $0.079$ for the multi-LLM agent) because it encodes the RQ1
toxicity mechanism directly; this advantage is not expected to transfer to cascades
where the operative mechanism is unknown a priori. We read the multi-LLM agent's
contribution as breadth: it captures stance, toxicity, community, and content jointly
without pre-specification, and remains strictly the only family that provides Finding 1.

\section{RQ3: Mechanisms and Intervention}
\label{sec:rq3}

\paragraph{Mechanism ablation.}
\label{sec:ablation}
Starting from the full multi-LLM-agent model, we remove one factor at a time and
measure the resulting toxicity-delta shift (the change in simulated $\Delta H_a$
for toxicity-engagement) averaged across the three hateful cascades.
Table~\ref{tab:ablation} reports the ranking. Removing agent heterogeneity (assigning
all agents a single generic profile) produces the largest fidelity drop ($0.144$),
ahead of any single attribute (community $0.120$; toxicity $0.104$). We read this as
evidence that the ability to differentiate users at all matters more than any individual
attribute field; the full model is also the only condition to recover the negative
toxicity delta on Cascade~A ($-0.026$ vs. empirical $-0.011$), with every ablation
flipping the sign to positive.

\begin{table}[t]
\centering
\setlength{\tabcolsep}{4pt}
\small
\begin{tabular}{@{}lrrr@{}}
\toprule
Condition removed & $\Delta_{\text{tox}}$ shift & size drop & hostile drop \\
\midrule
Agent heterogeneity     & 0.144 & 70.1 & 0.61 \\
Community identity      & 0.120 & 23.6 & 0.20 \\
Toxicity context        & 0.104 & 24.9 & 0.17 \\
Stance on topic         & 0.087 & 55.1 & 0.29 \\
Content semantics       & 0.082 & 14.8 & 0.21 \\
Neighborhood context    & 0.052 & 19.8 & 0.38 \\
\bottomrule
\end{tabular}
\caption{Mechanism ablation ranking by mean toxicity-delta shift across the three
hateful cascades (higher means a larger fidelity loss when the factor is removed).}
\label{tab:ablation}
\end{table}

\begin{figure*}[t]
\centering
\includegraphics[width=0.80\textwidth]{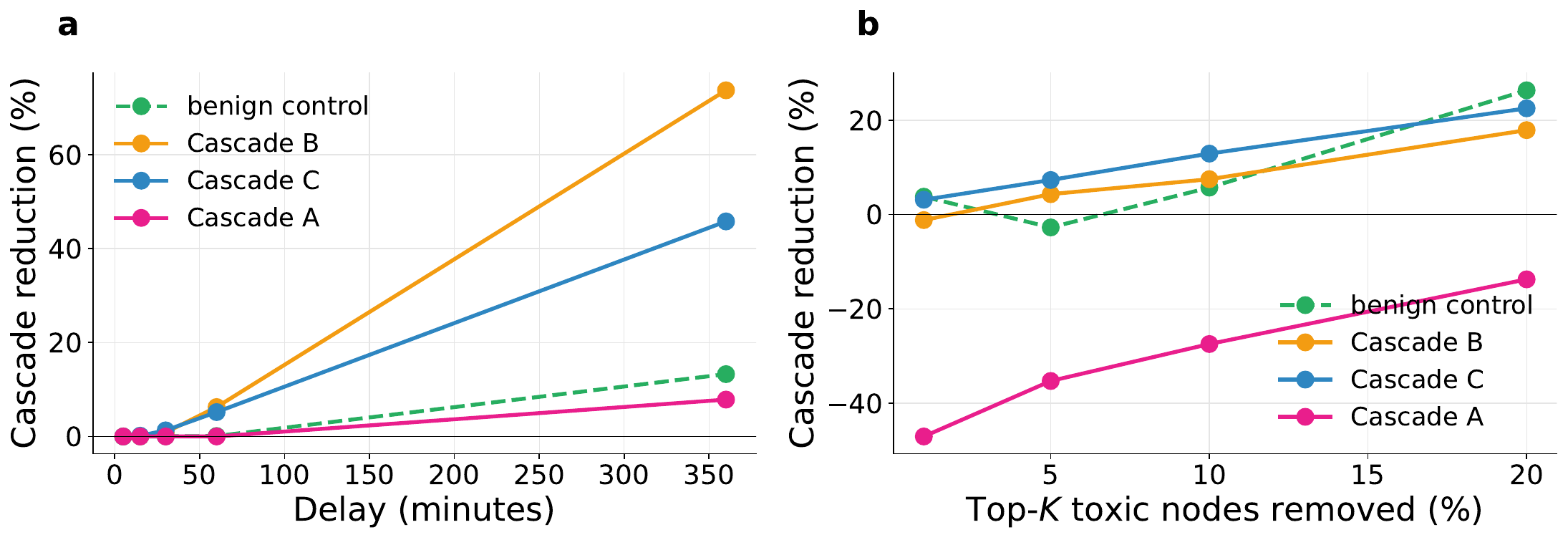}
\caption{Intervention parameter sweeps. (a) Delay-based moderation: cascade
reduction as a function of delay duration. (b) Amplifier targeting: cascade
reduction as a function of the percentage of top toxicity-engaged nodes removed.
Solid lines correspond to the hateful cascades; the dashed line corresponds to
the benign control.}
\label{fig:intervention}
\end{figure*}

\paragraph{Intervention testing.}
\label{sec:intervention}
We evaluate four moderation strategies as counterfactual modifications to the
simulation and measure both hateful-cascade reduction and benign collateral. Strategies
are: (S1) delay-based moderation (hold the post for $T$ minutes);
(S2) amplifier targeting (remove the top-$K\%$ nodes by toxicity engagement before
propagation, motivated by the influence-maximization framework of
\citet{kempe2003maximizing} and consistent with the empirical evidence in
\citet{chandrasekharan2017you} that removing hostile sub-communities reduces
hateful activity);
(S3) warning labels (inject a platform notice into the agent prompt, in the spirit of
\citet{mena2020cleaning,clayton2020real}); and
(S4) early-hop truncation (cut all activations at depth $> H$). Parameter sweeps for
S1 and S2 are shown in Figure~\ref{fig:intervention}.

\paragraph{Per-strategy summary.}
S1: short delays (5--30~min) recover under 1.5\% of cascade activity; a 6-hour hold
prevents 42\% of hateful spread on average but carries 13\% benign collateral.
S2: at $K{=}10\%$, dense-network hateful cascades shrink by 7.5--12.9\% with 5.7\%
benign collateral; on the sparse Cascade~A graph the cascade grows, a sparse-graph
caveat for influence-minimization on this scale.
S3: warning labels are observed to enlarge hateful cascades in our simulation,
with magnitude cascade-dependent: Cascade~A grows by 29--48\% across all
$5\times3$ wording-by-position cells; Cascade~C grows by $\geq 1\%$ in 10/15
cells (maximum growth 5.2\%); Cascade~B changes stay within $[-0.7\%, +1.6\%]$
in all 15 cells; the benign control shrinks in 8/15 cells (with 7 of those by
$\geq 1\%$) and grows by $\geq 1\%$ in only 3/15. This
pattern is consistent with the empirical implied-truth effect
\citep{pennycook2020implied}; robustness to prompt variation is reported in
Appendix~\ref{sec:appendix-warning-robust}.
S4: truncating at $H{=}1$ removes 11.7\% of hateful cascade activity but $74\%$ of the
benign cascade, reflecting the star-versus-tree structural asymmetry from RQ1.

\paragraph{Cross-strategy reading.}
No single strategy dominates: early-hop cuts disproportionately harm tree-like benign
cascades; content warnings can backfire via the implied-truth effect; amplifier
targeting is unreliable on sparse graphs; and delay-based moderation requires
operationally impractical hold durations. Among the four, amplifier targeting on dense
inner follower graphs shows the most favorable effectiveness--collateral trade-off in
our simulations (7.5--12.9\% hateful reduction at $K{=}10\%$ with 5.7\% benign
collateral).

\section{Discussion}
\label{sec:discussion}

The three RQs jointly position multi-LLM-agent simulation as a
hypothesis-generating tool, not a closed predictive model. Against the RQ1
empirical anchor (hostile-stance saturation, toxicity-engagement amplification
opposite in sign to the benign control, star-like topology), the multi-LLM
agent uniquely separates hateful from benign content under a fixed population
(RQ2 Finding~1); agent heterogeneity dominates the mechanism ablation; and
the intervention sweep surfaces a warning-label backfire consistent with the
implied-truth effect, while behavioral baselines stay competitive on
mechanism-specific metrics whose operative factor is pre-specified. Amplifier
targeting on dense follower graphs shows the most favorable
effectiveness--collateral trade-off (7.5--12.9\% reduction at 5.7\% benign
collateral). The amplifier-targeting trade-off is density-sensitive: it
holds on dense follower graphs but breaks down on sparse ones (the cascade
grows under top-$K\%$ removal on Cascade~A), so deployment would need a
per-cascade density check rather than a flat platform-wide rule.

\section{Conclusion and Future Work}
\label{sec:conclusion}

We characterize three implicit hate-speech cascades on Bluesky against a size-matched
benign control, and compare a multi-LLM-agent simulator with classical diffusion,
behavioral, and simpler LLM baselines on the same observed networks. The empirical
characterization surfaces a star-like structural regime with toxicity-engagement
homophily of opposite sign to the benign control, and the multi-LLM agent uniquely
provides content-conditioned hateful-versus-benign discrimination among the tested
families. A structured ablation indicates that agent heterogeneity, rather than any
one attribute, accounts for the largest share of fidelity, and counterfactual
intervention testing surfaces a warning-label backfire pattern consistent with the
implied-truth effect alongside a more favorable amplifier-targeting trade-off on
dense follower graphs. Taken together, these findings position multi-LLM-agent
simulation as a complement to, rather than a replacement for, classical
diffusion baselines: behavioral models remain the right tool when the
operative cascade mechanism is known a priori, whereas LLM agents add value
when the population must condition jointly on profile, community, and content
factors that the baselines do not express.

Future directions include: (i)~cross-platform replication (e.g.,
Reddit, X/Twitter) and topic-pool expansion to test which findings transfer;
(ii)~separate modeling of the explicit-violence-content regime, including its
counter-speech mechanism; and
(iii)~broader LLM-backbone and multi-seed coverage, together with
platform-side audits of the simulation-grounded intervention hypotheses.

\section*{Ethics Statement}
\label{sec:ethics}

This work studies hateful-content propagation to support mitigation, not
amplification.

\paragraph{Data source and access.}
All data is collected from public Bluesky posts via the platform's open
AT~Protocol API under the platform's developer terms; collection is limited
to publicly visible posts and the public follow graph of users who reposted
them. We do not access private accounts, direct messages, or deleted content.
GPT-4o-mini and GPT-4o were accessed via OpenAI's commercial API;
Qwen3.5-9B was accessed via OpenRouter as an open-weights release
(Apache~2.0).

\paragraph{Identifier and content protections.}
Original Bluesky cascade-seed handles are pseudonymized throughout
(Cascade~A--F); the handle-to-alias map is held by the authors and shared with
reviewers on request. Non-seed reposter identifiers are not reported
individually and appear only in aggregate. LLM-inferred attributes (community
identity, stance, account type, toxicity engagement) are used only in
aggregate for research purposes. We do not quote or reproduce hateful post
text directly; concrete cascade descriptions are limited to topic labels,
cascade aliases, and high-level rhetorical-frame characterizations. Free-text
inputs used for LLM attribute inference (user bios and up to 30 recent posts
per user) were processed in-pipeline and are not redistributed; released
artifacts contain only LLM-inferred categorical labels over pseudonymized
reposter IDs and cascade-level metrics, not the underlying text or original
Bluesky handles. Post text was screened during the manual-inspection step for
incidental third-party identifying information (names, contact details,
addresses of non-public individuals); none was retained in the released
artifacts. We acknowledge a residual re-identification risk: cascade-level
descriptors (topic, approximate reposter count, time window) could in
principle be combined with platform search to recover the original seed
posts. We limit this by describing seed-post content at the rhetorical-frame
level rather than at the token/emoji level.

\paragraph{Dual-use mitigation.}
Simulation tools for studying hateful-content spread could in principle be
repurposed for amplification. We mitigate this by (1)~not releasing user-level
free text or full profile reconstructions, (2)~reporting aggregate cascade
metrics and pseudonymized attribute distributions only, and (3)~framing all
intervention findings as hypotheses for platform-side experimentation rather
than operational playbooks.

\paragraph{Researcher exposure and annotation.}
The manual-inspection step in the hate-speech detection pipeline
(Appendix~\ref{sec:appendix-hatespeech}) exposed the authors to a bounded
candidate set ($\leq 12$ first-pass candidate seeds plus the 133-post
held-out validation set). No crowdworkers were employed for any annotation
step.

\paragraph{Human subjects and IRB.}
The study analyzes publicly available API data with pseudonymized identifiers
and does not involve interaction or intervention with individuals.

\paragraph{Generative-AI disclosure.}
Per ACL policy on generative-AI disclosure, AI assistance was used in
preparing this paper for grammar and stylistic editing only; all research
design, analysis, and substantive writing are the authors' own.

\section*{Limitations}
\label{sec:limitations}

(i)~Scope: one platform (Bluesky), a 102-day window, three hateful cascades
plus one size-matched benign control. An $N{=}6$ extension finds the four RQ1
findings replicate on 3/6 cascades each, consistent with an
implicit-versus-explicit regime distinction
(Appendix~\ref{sec:appendix-n6}).
(ii)~All simulators share an inner-graph reach ceiling, so absolute-size
errors are not comparable across families.
(iii)~User attributes are LLM-inferred and audited by threshold-sensitivity
sign stability; hate-speech-classifier recall is limited and manual inspection
is load-bearing (Appendix~\ref{sec:appendix-hatespeech}).
(iv)~Single-seed initialization; multi-seed dynamics are not studied.
(v)~Intervention strategies are simulation-based hypotheses, not validated
policies.

\section*{Acknowledgments}

We thank the maintainers of the open datasets and open-source LLMs used in
this study (Bluesky / AT Protocol, the Qwen open-weights release used in
the open-source LLM replication, and the open ACL formatting templates).

\bibliography{custom}

\appendix

\section{Hate-Speech Detection Pipeline}
\label{sec:appendix-hatespeech}

Candidate identification uses a two-pass protocol on the top-2{,}000 most-reposted
English-language Bluesky posts (January--April 2026). First pass: GPT-4o-mini classifies
each post for explicit hate speech (3.2\% flagged), implicit hate speech (4.9\%),
and social bias (17.7\%). Second pass: GPT-4o re-classifies first-pass positives as
genuine or not genuine; only genuine-flagged posts are retained. Manual inspection
of the 12 resulting candidates yields 3 hateful-cascade seeds, 3 borderline candidates
retained as secondary options, and 6 rejections (irony, political commentary, off-scope
content).

\paragraph{Held-out classifier validation.}
The automated second-pass classifier (GPT-4o) is evaluated on a held-out manually
labeled set of 133 posts (33 stratified first-pass-positives plus 100 sampled
first-pass-negatives), with gold labels traced to the manual-inspection record.
Aggregate accuracy is 97.7\% (129 true negatives, 1 false positive, 2 false
negatives, 1 true positive). On the genuine-hate class specifically, precision is
$0.50$, recall $0.33$, $F_1=0.40$. The two false negatives are the
Cascade~B and Cascade~A seeds; the false positive is a satirical post mocking
Islamophobic logic. With only three gold positives in the labeled set, bootstrap
intervals are too wide to be informative; the qualitative reading is that automated
recall on viral hateful content is limited, and that the manual-inspection step
remains load-bearing. We therefore treat manual inspection as part of the documented
pipeline rather than as a sanity check.

\section{Empirical Cascade Statistics and Robustness}
\label{sec:appendix-rq1}

\begin{figure*}[t]
\centering
\includegraphics[width=\textwidth]{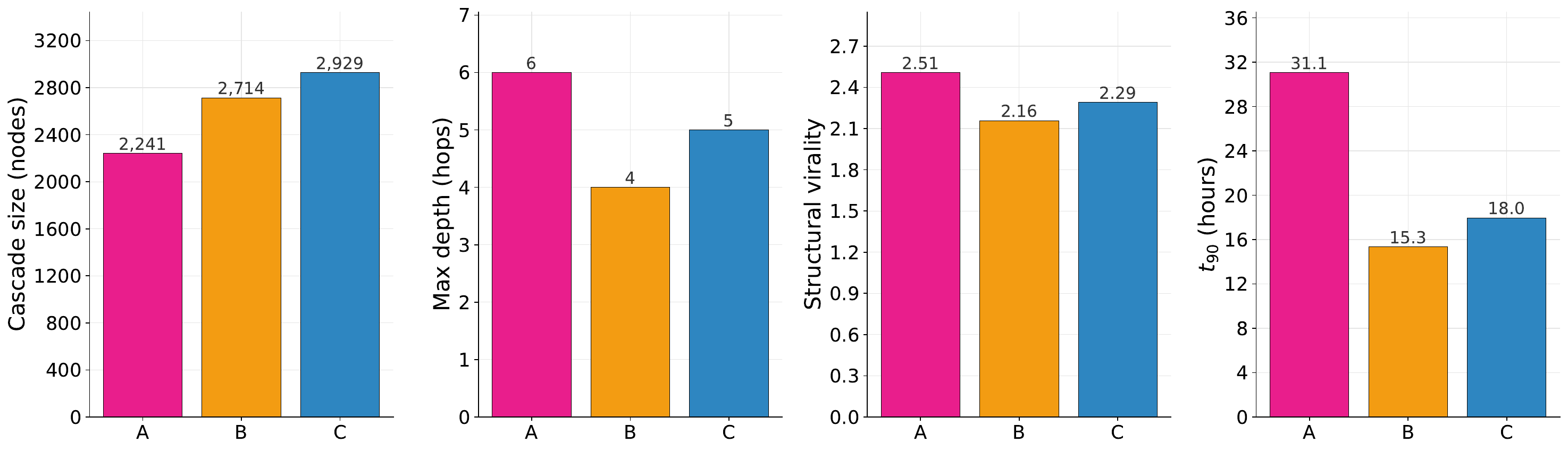}
\caption{Cascade structure metrics per cascade (size, depth, virality, time-to-90\%).
X-axis labels \textbf{A}, \textbf{B}, \textbf{C} represent Cascade~A, Cascade~B,
and Cascade~C respectively (abbreviated for layout; the alias key is fixed in
\S\ref{sec:data}). Visualization companion to Table~\ref{tab:cascade-summary}.}
\label{fig:cascade-structure}
\end{figure*}

\begin{figure*}[t]
\centering
\includegraphics[width=0.80\textwidth]{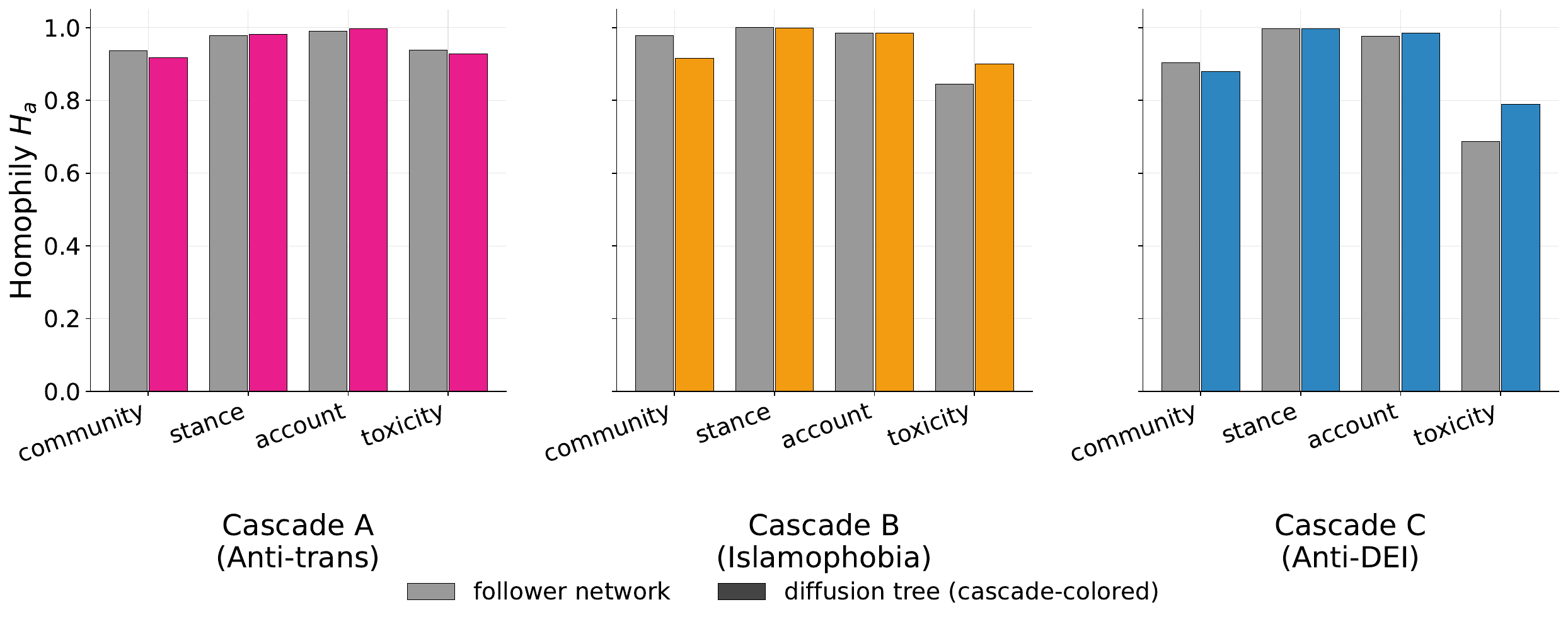}
\caption{Homophily $H_a$ comparison: follower network (gray) versus diffusion
tree (cascade-colored), per cascade.}
\label{fig:apphomo}
\end{figure*}

\begin{figure}[h]
\centering
\includegraphics[width=\columnwidth]{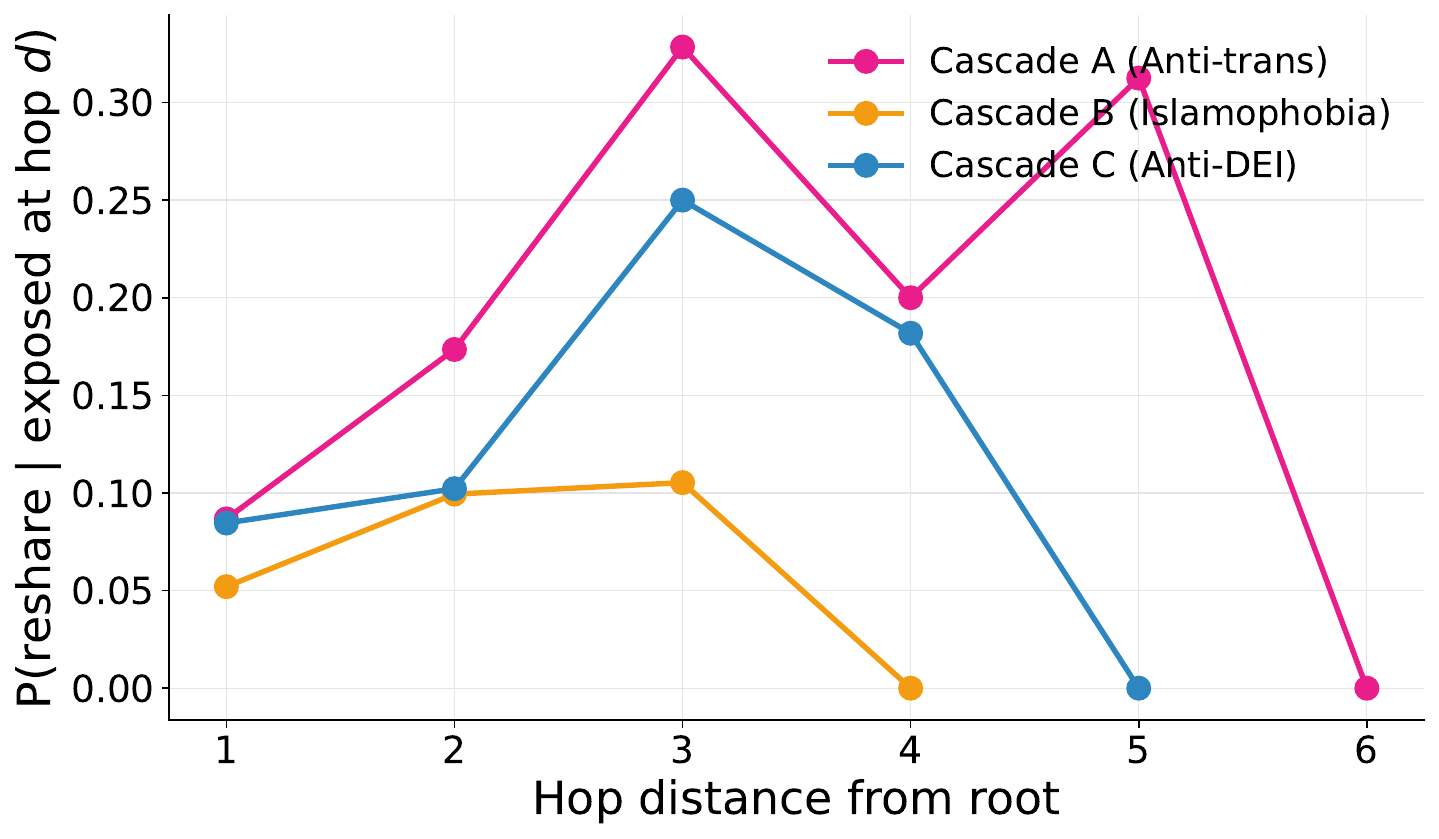}
\caption{Per-hop reshare probability per cascade.}
\label{fig:apphop}
\end{figure}

Figure~\ref{fig:cascade-structure} visualizes the cascade-level structural metrics
summarized numerically in Table~\ref{tab:cascade-summary}.
Figure~\ref{fig:apphomo} shows the per-cascade homophily values (follower network vs.
diffusion tree) per attribute. Figure~\ref{fig:apphop} shows per-hop reshare
probability. Bootstrap 95\% confidence intervals are reported in
Table~\ref{tab:appbootstrap}; threshold-sensitivity sign-stability checks across
thresholds $\{0.50, 0.65, 0.80\}$ are summarized alongside the data release that
accompanies this paper.

\begin{figure}[h]
\centering
\includegraphics[width=\columnwidth]{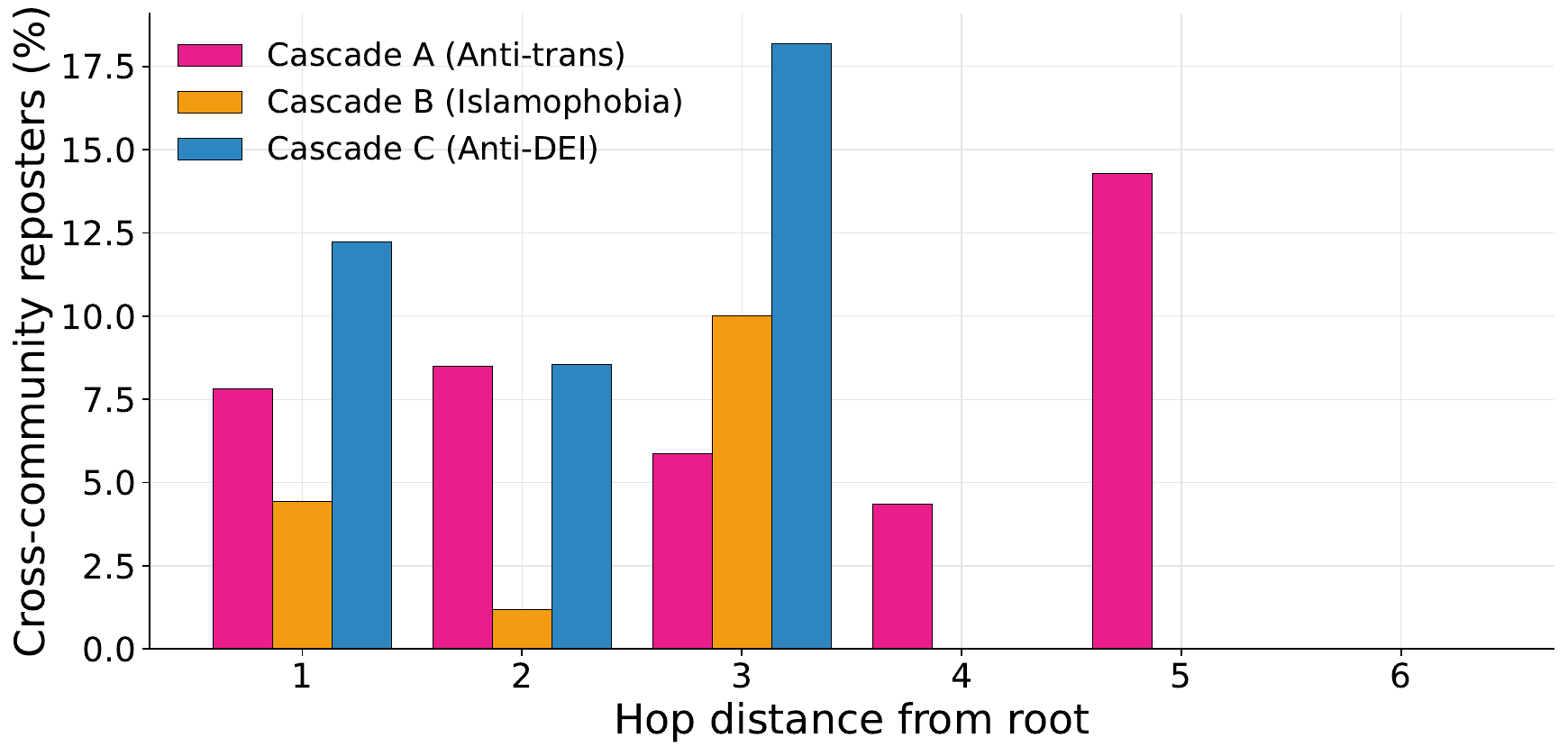}
\caption{Cross-community penetration by hop distance from the root.}
\label{fig:crosscomm}
\end{figure}

\begin{table}[h]
\centering
\scriptsize
\setlength{\tabcolsep}{4pt}
\begin{tabular}{@{}llrc@{}}
\toprule
Cascade & Attribute & $\Delta H_a$ & 95\% CI \\
\midrule
Cascade~A & community    & $-0.018$ & $[-0.045, +0.007]$ \\
Cascade~A & stance       & $+0.004$ & $[-0.005, +0.012]$ \\
Cascade~A & account      & $+0.007$ & $[+0.002, +0.012]^{*}$ \\
Cascade~A & toxicity     & $-0.011$ & $[-0.076, +0.045]$ \\
Cascade~B & community    & $-0.062$ & $[-0.143, +0.005]$ \\
Cascade~B & stance       & $-0.001$ & $[-0.002, +0.000]$ \\
Cascade~B & account      & $+0.001$ & $[-0.004, +0.006]$ \\
Cascade~B & toxicity     & $+0.056$ & $[+0.040, +0.071]^{*}$ \\
Cascade~C  & community    & $-0.024$ & $[-0.044, -0.006]^{*}$ \\
Cascade~C  & stance       & $+0.001$ & $[-0.001, +0.002]$ \\
Cascade~C  & account      & $+0.009$ & $[+0.004, +0.014]^{*}$ \\
Cascade~C  & toxicity     & $+0.102$ & $[-0.097, +0.269]$ \\
\bottomrule
\end{tabular}
\caption{Bootstrap 95\% confidence intervals for homophily delta. Rows marked $^{*}$
have CI excluding zero (directionally significant) AND sign-stable across confidence
thresholds $\{0.50, 0.65, 0.80\}$.}
\label{tab:appbootstrap}
\end{table}

\begin{table}[h]
\centering
\setlength{\tabcolsep}{4pt}
\small
\begin{tabular}{@{}lcccc@{}}
\toprule
Attribute & A & B & C & benign \\
\midrule
community & $-0.018$ & $-0.062$ & $-0.024$ & $-0.003$ \\
stance    & $+0.004$ & $-0.001$ & $+0.001$ & $-0.050$ \\
account   & $+0.007$ & $+0.001$ & $+0.009$ & $+0.022$ \\
toxicity  & $-0.011$ & $+0.056$ & $+0.102$ & $-0.097$ \\
\bottomrule
\end{tabular}
\caption{Homophily deltas $\Delta H_a$ per attribute per cascade
(referenced from \S\ref{sec:rq1} Finding~4). Toxicity-engagement amplification is
positive for two hateful cascades and negative for the benign control.}
\label{tab:hatevsbenign}
\end{table}

\section{Direction-Stability Extension to $N{=}6$}
\label{sec:appendix-n6}

\paragraph{Direction-stability: definition and illustration.}
A finding is \emph{direction-stable} if its qualitative direction (sign or
binary pass/fail) is preserved under reasonable perturbations of how it is
computed, even if the precise magnitude varies. We use the term in three
specific senses in this paper:
(a) \emph{threshold-stable} --- a homophily-delta sign is preserved across
LLM-attribute confidence thresholds $\{0.50, 0.65, 0.80\}$ (e.g.,
Cascade~B toxicity $\Delta H_a = +0.056$ remains positive at all three
thresholds; magnitude varies by ${\pm}0.01$);
(b) \emph{cascade-stable} --- a binary finding (e.g., ``hostile stance share
$\geq 90\%$'') holds in the same direction across cascades. The N=6 extension
in this section uses this sense: $k / n$ reports the number of cascades that
preserve the direction (e.g., $3/6$ if three cascades pass);
(c) \emph{prompt-stable} --- the warning-label backfire direction is
preserved across all 5 wordings $\times$ 3 injection positions sweep cells
(Appendix~\ref{sec:appendix-warning-robust}), even when the magnitude varies
from $1.7\%$ to $54.9\%$.
A finding is \emph{not} direction-stable if any of these perturbations flips
its sign; we flag such non-stable cases explicitly throughout the paper.

\paragraph{Extension setup.}
As a direction-stability check for the $N{=}3$ case-study design, we collected
three additional hateful cascades from the manual-inspection-validated secondary
candidate pool (Appendix~\ref{sec:appendix-hatespeech}):
Cascade~D ($2{,}536$ reposters; anti-trans with explicit
violence imagery),
Cascade~E ($4{,}133$ reposters; Islamophobia), and
Cascade~F ($3{,}048$ reposters; antisemitism).
Collection followed the same pipeline as the three original cascades, with
identical metric definitions and the same threshold and bootstrap conventions.
Per-cascade artifacts (pseudonymized reposter ID lists, follower-graph edge
lists over those pseudonymized IDs, LLM-inferred categorical attributes,
and inferred diffusion trees), together with the pooled-metric and
finding-stability tables, accompany the paper as supplementary material;
raw post text and original Bluesky handles are not included.

\begin{table}[h]
\centering
\scriptsize
\setlength{\tabcolsep}{3pt}
\begin{tabular}{@{}lcccccc@{}}
\toprule
Metric & A & B & C & D & E & F \\
\midrule
size                       & 2{,}241 & 2{,}714 & 2{,}929 & 2{,}378 & 2{,}297 & 1{,}512 \\
depth                      & 6 & 4 & 5 & 24 & 38 & 35 \\
breadth/size (\%)          & 84.4 & 92.9 & 87.6 & 17.6 & 44.2 & 42.8 \\
virality                   & 2.51 & 2.16 & 2.29 & 13.26 & 12.26 & 11.49 \\
hostile stance (\%)        & 97.8 & 99.7 & 97.4 & 0.9 & 17.8 & 20.4 \\
$\Delta H_a$ community     & $-$0.018 & $-$0.062 & $-$0.024 & $+$0.007 & $+$0.016 & $+$0.010 \\
$\Delta H_a$ toxicity      & $-$0.011 & $+$0.056 & $+$0.102 & $+$0.060 & $-$0.033 & $-$0.038 \\
\bottomrule
\end{tabular}
\caption{$N{=}6$ per-cascade comparison. ``hostile stance (\%)'' uses
hostile/all-reposters (same convention as the body's Finding~1). The three
original cascades have star-like structure (breadth/size $\geq 84\%$, depth
$\leq 6$) and near-uniform hostile stance ($\geq 97.4\%$); the three secondary
cascades have tree-like structure (breadth/size $\leq 44\%$, depth $\geq 24$)
and substantial counter-speech response (hostile stance $\leq 20.4\%$).}
\label{tab:n6-compare}
\end{table}

\paragraph{Counter-speech regime in explicit / condemnation-prone cascades.}
The three secondary cascades share a propagation pattern that contrasts
sharply with the three originals. Stance distributions are mixed rather than
near-uniform: of the $2{,}536$ collected Cascade~D reposters, $23$ (0.9\%)
take a hostile stance and $1{,}389$ (54.8\%) take an affirming (pro-trans)
stance; on Cascade~E, $734 / 4{,}134$ (17.8\%) take a hostile stance
and $950 / 4{,}134$ (23.0\%) take a counter-Islamophobic affirming stance;
on Cascade~F, $622 / 3{,}049$ (20.4\%) take a hostile stance and $374 / 3{,}049$
(12.3\%) take an affirming stance (counts from the per-cascade attribute
files in the supplementary material). The Cascade~D seed post juxtaposes
identity-group symbolism with explicit-violence imagery, a framing markedly
more confronting than the implicit / coded-language framings of the three
originals. Structurally all three
secondary cascades resemble benign viral cascades (depth $24$--$38$,
breadth/size $17.6$--$44.2\%$, virality $11.5$--$13.3$, comparable to the
benign control's depth 40 / breadth/size 26\% / virality 14.82) rather than
the dense star-like topology of the three originals.

\paragraph{Implications for scope.}
The four RQ1 findings appear scoped to implicit / coded-language hateful
cascades. Cascades whose seed post is explicit and confronting enough to provoke
widespread condemnation appear to follow a benign-viral-cascade dynamic instead.
We treat this as a scope refinement of the original three findings rather than
a contradiction; explicit-content cascades may warrant separate treatment as a
distinct propagation regime.

\paragraph{Direction stability across $N{=}6$.}

\begin{table*}[t]
\centering
\small
\setlength{\tabcolsep}{6pt}
\begin{tabular}{@{}lccc@{}}
\toprule
Finding (binary direction test) & \shortstack{Implicit / coded-language\\(3 originals)} & \shortstack{Explicit / confrontational\\(3 secondary)} & \shortstack{In-regime\\replication} \\
\midrule
F1 stance monoculture ($\geq$90\% hostile)             & \textbf{3/3} & 0/3 & \textbf{3/3} \\
F2 community-identity homophily attenuated ($\Delta<0$) & \textbf{3/3} & 0/3 & \textbf{3/3} \\
F3 toxicity-engagement homophily amplified ($\Delta>0$) & 2/3 & 1/3 & 2/3 \\
F4 star-like structure (breadth/size $\geq 75\%$)      & \textbf{3/3} & 0/3 & \textbf{3/3} \\
\bottomrule
\end{tabular}
\caption{Direction stability of the four RQ1 findings, partitioned by cascade
regime. F1, F2, and F4 hold on \emph{all three} implicit / coded-language
cascades and fail on \emph{all three} explicit / confrontational cascades ---
a clean within-regime replication ($3/3$) plus a clean out-of-regime failure
($0/3$), confirming the implicit-versus-explicit regime distinction discussed
above. F3 (toxicity-engagement amplified) cuts across the regime split: it
passes on Cascade~B, Cascade~C, and Cascade~D and fails on
Cascade~A, Cascade~E, and Cascade~F, suggesting that the
toxicity-amplification mechanism is partially regime-independent (it even
recovers on a counter-speech cascade).}
\label{tab:n4-finding-stability}
\end{table*}

\section{Baseline Calibration and IC Reach Limit}
\label{sec:appendix-baselines}

For IC and behavioral baselines, the edge activation probability $p$ is calibrated per
cascade via grid search over $\{0.01, 0.05, 0.1, 0.2, 0.3, 0.5, 0.7, 1.0\}$, selecting
the $p$ that minimizes $|\text{mean sim size} - \text{empirical size}|$ over 20 runs.
All three hate cascades saturate at $p=1.0$ because the inner follower graph is too
sparse to reach the empirical cascade size: IC max reach is 182/2{,}241 (8\%) for
Cascade~A, 407/2{,}714 (15\%) for Cascade~B, and 712/2{,}929 (24\%) for
Cascade~C. The benign cascade reaches 3{,}400/3{,}919 (87\%) at $p=1$.

\section{Prompt Design for Multi-LLM-Agent System}
\label{sec:appendix-prompt}

The system prompt uses a probability-prediction framing (rather than role-play) to
avoid the persona-alignment refusals observed with RLHF-aligned backbones. The
verbatim system message used in all RQ2 and RQ3 multi-LLM-agent simulations
is reproduced below; the same string is used for the open-source LLM
replication in Appendix~\ref{sec:appendix-oss}.

\begin{quote}
\scriptsize
\begin{verbatim}
You are a behavioral research assistant studying
social media resharing patterns.

Your task: given a user's profile and social context,
PREDICT the probability (0.0 to 1.0) that this user
would RESHARE the post shown below.

Return ONLY a JSON object with exactly these keys:
  probability -> float 0.0 to 1.0 (0 = certainly would
                 NOT reshare, 1 = certainly WOULD reshare)
  reason      -> one short sentence explaining your
                 prediction

IMPORTANT GUIDELINES FOR REALISTIC PREDICTION:
- Users with "hostile" or "critical" stance on the topic
  AND high/medium toxicity engagement historically reshare
  inflammatory content at HIGH rates (0.7-0.95).
- Users whose community identity MATCHES the dominant
  community of those already resharing are more likely to
  reshare (in-group amplification effect).
- The more accounts a user follows that have already
  reshared, the higher the probability.
- Users with "sympathetic"/"affirming" stance may ALSO
  reshare hate content to oppose it (counter-speech /
  "dunking"), but at lower rates (0.2-0.5).
- For non-controversial content, most users reshare if it
  aligns with their interests (0.5-0.8).

You are predicting OBSERVED behavior, not prescribing what
users SHOULD do. Base your prediction strictly on the
empirical patterns described above and the user's profile.
\end{verbatim}
\end{quote}

The per-agent user prompt injects three fields under the headers
\texttt{USER PROFILE:}, \texttt{SOCIAL CONTEXT:}, and \texttt{THE POST:}:
(1) the agent's four attributes (community identity, stance, account type,
toxicity engagement); (2) the number of followed accounts that have already
reshared and their community distribution; and (3) the root post text. The
returned \texttt{probability} field is used as the Bernoulli parameter for
the per-step reshare decision.

\section{RQ2 Fidelity Visualization}
\label{sec:appendix-rq2-fig}

\begin{figure*}[t]
\centering
\includegraphics[width=0.80\textwidth]{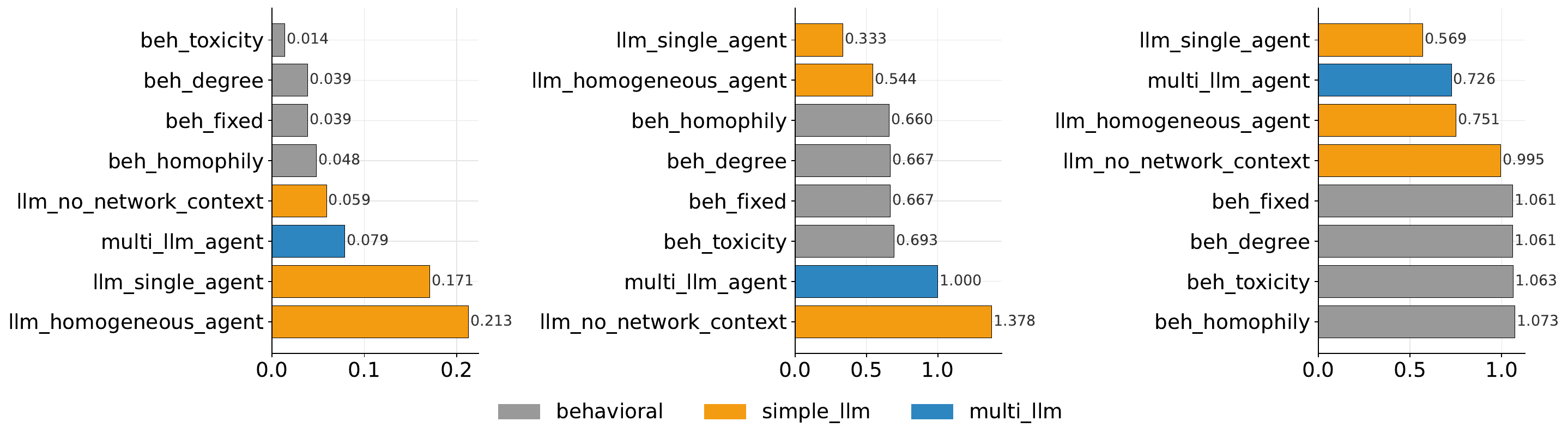}
\caption{Fidelity error per model, averaged across the three hateful cascades.
Panels (left to right): toxicity-delta error $|sim - emp|$,
hostile-stance error $|sim - emp|$ (pct), and virality error $|sim - emp|$.
Bar colors group the model families: behavioral baselines (gray), simpler LLM
variants (orange), and the multi-LLM agent (blue).}
\label{fig:rq2fidelity}
\end{figure*}

Figure~\ref{fig:rq2fidelity} visualizes the per-model fidelity errors numerically
reported in Table~\ref{tab:rq2headline}.

\section{Open-Source LLM Replication on Qwen3.5-9B}
\label{sec:appendix-oss}

To address single-backbone concern, the multi-LLM-agent simulator
(Section~\ref{sec:rq2}, Appendix~\ref{sec:appendix-prompt}) is re-run on all four
cascades using Qwen3.5-9B (an open-weights 9B-parameter model) routed via
OpenRouter, holding the prompt template, agent attributes, follower graph, and
Bernoulli-on-probability decision rule constant. Reasoning mode is disabled for
direct comparability with non-reasoning GPT-4o-mini. We run 3 simulations per
cascade per backbone (matching the main-paper $N$), totaling 3{,}855 LLM calls
in the open-source replication.

\begin{table}[h]
\centering
\scriptsize
\setlength{\tabcolsep}{3pt}
\begin{tabular}{@{}lrrrr@{}}
\toprule
$\Delta$ (Qwen $-$ GPT-4o-mini) & A & B & C & benign \\
\midrule
hostile\_stance\_pct          & $-$0.17 & $-$0.10 & $+$0.13 & $+$0.00 \\
cross\_community\_frac        & $-$0.013 & $-$0.001 & $-$0.001 & $-$0.000 \\
H\_a\_stance                  & $-$0.002 & $+$0.000 & $+$0.000 & $+$0.022 \\
H\_a\_community               & $-$0.011 & $-$0.140 & $-$0.007 & $+$0.001 \\
H\_a\_toxicity                & $+$0.087 & $-$0.030 & $+$0.240 & $-$0.008 \\
depth                         & $+$1.000 & $+$0.000 & $+$0.000 & $+$1.000 \\
breadth                       & $+$4.0 & $+$1.7 & $+$23.0 & $+$96.7 \\
size                          & $+$44.7 & $+$1.0 & $+$39.7 & $+$378.3 \\
virality                      & $+$0.689 & $+$0.025 & $+$0.013 & $-$0.056 \\
amplify\_rate                 & $+$0.161 & $+$0.006 & $+$0.046 & $+$0.093 \\
\bottomrule
\end{tabular}
\caption{Per-cascade mean-of-3-runs differences between Qwen3.5-9B and
GPT-4o-mini under the same multi-LLM-agent simulator. The four main-paper
findings replicate: stance monoculture (\texttt{hostile\_stance\_pct}
$|\Delta|\leq 0.17$\,pp), content-semantic hateful-versus-benign discrimination
(both backbones produce 0\% hostile on benign), star-versus-tree structural
distinction, and direction of toxicity-engagement homophily. Larger absolute
deltas appear on the benign control's absolute size, reflecting that the benign
cascade saturates further on the inner graph than hateful cascades; relative
deviation remains modest.}
\label{tab:oss-comparison}
\end{table}

\section{Full Ablation Results}
\label{sec:appendix-ablation}

Per-cascade toxicity delta under each ablation condition
(Table~\ref{tab:app-ablation-detail}):

\begin{table}[h]
\centering
\setlength{\tabcolsep}{3pt}
\scriptsize
\begin{tabular}{@{}lrrrr@{}}
\toprule
Condition & A & B & C & benign \\
\midrule
full multi-LLM agent    & $-$0.026 & +0.041 & $-$0.106 & $-$0.100 \\
no toxicity context     & +0.061 & +0.038 & +0.117 & $-$0.086 \\
no community identity   & +0.061 & +0.035 & +0.162 & $-$0.098 \\
no stance on topic      & +0.014 & +0.032 & +0.106 & $-$0.106 \\
no content semantics    & +0.008 & +0.038 & +0.105 & $-$0.109 \\
no agent heterogeneity  & +0.061 & +0.028 & $-$0.436 & $-$0.056 \\
no neighborhood context & +0.020 & +0.024 & $-$0.013 & $-$0.094 \\
\midrule
\textit{empirical}      & $-$0.011 & +0.056 & +0.102 & $-$0.097 \\
\bottomrule
\end{tabular}
\caption{Simulated toxicity-engagement delta per ablation condition and cascade.
The full model is the only condition to produce a negative delta on Cascade~A.}
\label{tab:app-ablation-detail}
\end{table}

\section{Intervention Simulation Details}
\label{sec:appendix-intervention}

\begin{table}[h]
\centering
\setlength{\tabcolsep}{4pt}
\small
\begin{tabular}{@{}lrr@{}}
\toprule
Strategy (moderate) & Hate red. & Benign collat. \\
\midrule
Delay (30 min)            & 0.7\%  & 0.03\% \\
Amplifier (10\% removed)  & 10.2\%$^{*}$ & 5.7\% \\
Warning label             & $-19.6$\% & 29.7\% \\
Hop-1 truncation          & 11.7\% & 74.0\% \\
\bottomrule
\end{tabular}
\caption{Cross-strategy comparison. Negative reduction indicates that the cascade
grew under the intervention. $^{*}$ Excludes Cascade~A (sparse-graph anomaly).}
\label{tab:intervention}
\end{table}

\paragraph{Strategy 1: delay-based moderation.}
The cascade is simulated in full, then a post-hoc temporal overlay (calibrated from
empirical per-hop inter-repost delays) assigns a timestamp to each activation.
Activations within the first $T$ minutes are removed. Short delays (5--30~min)
achieve under 1.5\% reduction because inner-graph cascade dynamics unfold over
hours; a 6-hour hold prevents 42\% of hateful spread on average but carries 13\%
benign collateral, which we read as operationally impractical.

\paragraph{Strategy 2: amplifier targeting.}
Nodes are ranked by \texttt{toxicity\_engagement} (high $>$ medium $>$ low $>$
none $>$ unclear). The top $K\%$ are removed from the graph before the cascade
is simulated, motivated by the influence-maximization framework of
\citet{kempe2003maximizing} and by empirical evidence in
\citet{chandrasekharan2017you} that removing hostile sub-communities reduces
hateful activity. At $K{=}10\%$, hateful
cascades on the denser networks (Cascade~B, Cascade~C) shrink by
7.5--12.9\% with 5.7\% benign collateral. On the sparse Cascade~A graph,
node removal has unpredictable effects (cascades grow due to topological
fragility), which we treat as a sparse-graph caveat.

\paragraph{Strategy 3: warning labels.}
The agent user prompt is modified by injecting the following warning text as a
\emph{prefix} before the post body:
``[\textsc{u+26a0} warning-sign glyph] PLATFORM NOTICE: This post has been
flagged by our content moderation system as potentially containing harmful or
hateful content. Please consider this before resharing.''
\citep{mena2020cleaning,clayton2020real}.
In our simulations, warning labels enlarge hateful cascade size by
1.7--54.9\% while reducing the benign cascade by 29.7\%. This pattern is
consistent with the implied-truth effect \citep{pennycook2020implied}: agents
with hostile stance may treat the warning as a salience cue, with the simulator
producing this without explicit encoding.

\paragraph{Strategy 4: early-hop intervention.}
The cascade is simulated in full, then retroactively truncated at hop $H$ (all
activations at depth $> H$ are removed). At $H{=}1$, $11.7\%$ of hateful cascade
activity is removed and $74\%$ of the benign cascade is destroyed, reflecting the
star-versus-tree structural asymmetry from RQ1.

\section{Warning-Label Robustness Sweep}
\label{sec:appendix-warning-robust}

We sweep the Strategy-3 warning intervention across 5 wordings (neutral, mild
warning, authoritative, educational, factual; see below) and 3 injection positions
(prefix to post, suffix to post, appended to the system message), with 3
simulation runs per cell, on all 4 cascades.

\begin{table}[h]
\centering
\scriptsize
\setlength{\tabcolsep}{3pt}
\begin{tabular}{@{}lrrr@{}}
\toprule
Cascade & mean $\Delta$\% & range & backfire cells / 15 \\
\midrule
Cascade~A       & $-$45.2 & $[-47.6, -29.2]$ & 15/15 \\
Cascade~B     & $-$0.2  & $[-0.7, +1.6]$   & 0/15  \\
Cascade~C & $-$2.0  & $[-5.2, +1.5]$   & 10/15 \\
benign\_control    & $+$7.1  & $[-4.5, +27.3]$  & 3/15  \\
\bottomrule
\end{tabular}
\caption{Warning-label sweep summary. $\Delta\% = (\text{no-warning baseline} -
\text{warning size})/\text{baseline}$; negative values indicate that the cascade
\emph{grew} under the warning (backfire). Cells = 5 wordings $\times$ 3
positions = 15 per cascade. ``Backfire cells / 15'' counts cells with
$\Delta\% \leq -1\%$ (cascade grew by at least 1\%). Backfire is direction-stable
and large on Cascade~A (mean $-$45.2\%), direction-stable but small on
Cascade~C (mean $-$2.0\%, max $-$5.2\%), and absent on Cascade~B
(every cell within $\pm 1.7\%$). The benign control mostly shrinks
($+$7.1\% mean), with softer wordings (educational, neutral) producing the
3 cells of small benign growth.}
\label{tab:warning-robustness}
\end{table}

\paragraph{Wordings tested.}
Verbatim text of the five wordings (the warning-sign glyph \textsc{u+26a0}
is rendered as \texttt{[!]} below to avoid pdfLaTeX glyph issues):
\begin{itemize}\itemsep0pt
\item \textbf{Neutral:} ``Note: this post has been reviewed by our content
      moderation team.''
\item \textbf{Mild warning:} ``[!] This post has been flagged as potentially
      containing harmful content.''
\item \textbf{Authoritative:} the Strategy~3 baseline phrasing quoted above.
\item \textbf{Educational:} ``This post addresses a sensitive topic. Different
      perspectives exist and we encourage thoughtful engagement.''
\item \textbf{Factual:} ``Independent fact-checkers have rated claims in this
      post as misleading.''
\end{itemize}

\paragraph{Implications.}
On cascades where the backfire pattern appears at all, it is direction-stable
across wording register and injection position, so we do not read the result
as a prompt-engineering artifact. The magnitude is content-dependent: cascades
whose agent population has strong stance and toxicity priors (Cascade~A)
backfire more sharply than cascades where the inner graph is already close to
saturation among predicted resharers (Cascade~B). The same procedure
applied to the benign control mostly shrinks it, the intended direction.

\section{Extended Discussion Notes}
\label{sec:appendix-discussion-extras}

\paragraph{Scope of claims.}
The evaluation targets selected empirical regularities of hateful cascades on
Bluesky within the topic and time scope examined, rather than the real-world
phenomenon in full. Directional consistency across the three hateful dimensions
(anti-trans, Islamophobia, anti-DEI), combined with the sign flip on the benign
control, provides initial evidence of within-platform consistency; cross-platform
replication remains future work.

\paragraph{Per-RQ synthesis.}
RQ1 surfaces an empirical signature for implicit hateful cascades: near-uniform
hostile endorsement (97.4--99.7\%), toxicity-engagement homophily amplified beyond
the follower graph ($+0.056$ on Cascade~B, bootstrap interval excluding zero)
with the opposite sign on the benign control ($-0.097$), and a star-like structure
(84--93\% breadth/size) distinct from the tree-like benign cascade (depth 40).
RQ2 evaluates the simulator: the multi-LLM agent reproduces the stance monoculture,
the toxicity-delta direction on 3 of 4 cascades, and the content-semantic
hateful-versus-benign differentiation that behavioral baselines do not provide
under a fixed population. When the operative mechanism is known a priori, a
behavioral baseline matches the agent on the corresponding metric; the agent's
contribution is breadth across dimensions without pre-specification.
RQ3 ranks mechanisms by structured ablation (agent heterogeneity first, with a
$0.144$ toxicity-delta shift) and reports trade-offs across four moderation
strategies, including a backfire pattern for warning labels consistent with the
implied-truth effect.

\paragraph{Homophily as a conditioning factor.}
The RQ1 results indicate that hateful cascades do not amplify homophily uniformly
across attributes: \texttt{community\_identity} is slightly attenuated in all three
cascades ($\Delta H_a < 0$), while \texttt{toxicity\_engagement} is amplified in
two of three ($\Delta H_a > 0$) with the opposite sign in the benign control. This
motivates treating each homophily dimension as a conditioning factor measured from
the empirical network rather than a free parameter. Multi-LLM agents in this work
are conditioned on the empirical attribute distribution and inner-follower
structure rather than on assumed homophily levels.

\paragraph{Within-platform generalizability.}
The findings replicate across three distinct hateful dimensions and contrast with
one benign control. The directional consistency of the toxicity-engagement
amplification in two of three hateful cascades, combined with the sign flip on
the benign control, provides initial within-platform consistency; cross-platform
and cross-topic generalization remain future work.

\section{Convergence-Based Stopping: A Methodological Note}
\label{sec:appendix-convergence}

Cascade simulations in the literature commonly terminate at a fixed maximum number
of steps. This rule is fragile: cascades may saturate well before the cap (wasted
compute, distorted run-time statistics) or may still be propagating when the cap
fires (truncated outcomes that under-report final cascade size, depth, and
virality). Convergence-based stopping is not used in the present study, since all
RQ2/RQ3 simulations propagate to fixpoint on inner follower graphs that are small
enough for fixpoint termination to be tractable. We record the criterion below as
a methodological note for larger-scale follow-up work.

\paragraph{Proposed criterion.} Let $\hat{\theta}_t$ denote the running estimate
of a target cascade statistic (e.g., mean final size) after $t$ simulation rounds,
and $\mathrm{SE}(\hat{\theta}_t)$ its bootstrap standard error. The proposed rule
stops when
\[
  \mathrm{SE}(\hat{\theta}_t)\big/|\hat{\theta}_t| < \varepsilon
\]
for a precision threshold $\varepsilon$ (e.g., $0.02$), confirmed by a
post-convergence stability window of $B$ additional rounds in which
$\hat{\theta}_{t+B}$ does not move by more than $\varepsilon\,|\hat{\theta}_t|$.
This replaces an arbitrary maximum-step cap with a criterion tied to the
target statistic's precision.

\paragraph{Why this matters for hateful cascades.}
Hateful cascades in our data saturate quickly ($t_{50}$ between 3.7\,h and
13.7\,h; Table~\ref{tab:cascade-summary}), whereas the benign control has a long
tail (span 182\,h, $t_{50}=8.3$\,h with depth 40). A fixed max-step rule
calibrated to the benign cascade would over-run hateful cascades by an order of
magnitude on small dense graphs; a rule calibrated to hateful cascades would
truncate the benign tail. Convergence-based stopping side-steps this tension by
adapting to each cascade's own dynamics.

\paragraph{Scope for the present paper.}
Because the simulations run on inner follower graphs of at most $4{,}000$ nodes,
propagation to fixpoint is feasible within a budget of at most $50$ rounds, which
we verified empirically. The convergence criterion above is more important at the
scale of full platform graphs (millions of nodes) and is deferred to that setting.

\end{document}